\documentclass[aps,prl,superscriptaddress,reprint,sort&compress]{revtex4-1}
\usepackage[latin9]{inputenc}
\usepackage{amsmath}
\usepackage{graphicx}

\makeatletter

\providecommand{\tabularnewline}{\\}

\makeatother

\begin{document}

\title{Optical probing of mechanical loss of a Si$_{3}$N$_{4}$ membrane
below 100 mK}

\author{R. Fischer}

\email{ran.fischer@jila.colorado.edu}

\affiliation{JILA, National Institute of Standards and Technology and University
of Colorado, and Department of Physics, University of Colorado, Boulder,
Colorado 80309, USA}

\author{N. S. Kampel}

\affiliation{JILA, National Institute of Standards and Technology and University
of Colorado, and Department of Physics, University of Colorado, Boulder,
Colorado 80309, USA}

\author{G. G. T. Assumpç$\tilde{\textrm{a}}$o}

\affiliation{JILA, National Institute of Standards and Technology and University
of Colorado, and Department of Physics, University of Colorado, Boulder,
Colorado 80309, USA}

\author{P.-L. Yu}

\thanks{Current address: School of Electrical and Computer Engineering, Birck
Nanotechnology Center, Purdue University, West Lafayette, IN 47907
USA}

\affiliation{JILA, National Institute of Standards and Technology and University
of Colorado, and Department of Physics, University of Colorado, Boulder,
Colorado 80309, USA}

\author{K. Cicak}

\affiliation{National Institute of Standards and Technology, 325 Broadway, Boulder,
Colorado 80305, USA}

\author{R. W. Peterson}

\affiliation{JILA, National Institute of Standards and Technology and University
of Colorado, and Department of Physics, University of Colorado, Boulder,
Colorado 80309, USA}

\author{R. W. Simmonds}

\affiliation{National Institute of Standards and Technology, 325 Broadway, Boulder,
Colorado 80305, USA}

\author{C. A. Regal}

\email{regal@colorado.edu}

\affiliation{JILA, National Institute of Standards and Technology and University
of Colorado, and Department of Physics, University of Colorado, Boulder,
Colorado 80309, USA}
\begin{abstract}
We report on low mechanical loss in a high-stress silicon nitride
(Si$_{3}$N$_{4}$) membrane at temperatures below 100 mK. We isolate
a membrane via a phononic shield formed within a supporting silicon
frame, and measure the mechanical quality factor of a number of high-tension
membrane modes as we vary our dilution refrigerator base temperature
between 35 mK and 5 K. At the lowest temperatures, we obtain a maximum
quality factor ($Q$) of $2.3\times10^{8}$, corresponding to a $Q$-frequency
product (QFP) of $3.7\times10^{14}$ Hz. These measurements complement
the recent observation of improved quality factors of Si$_{3}$N$_{4}$
at ultralow temperatures via electrical detection. We also observe
a dependence of the quality factor on optical heating of the device.
By combining exceptional material properties, high tension, advanced
isolation and clamping techniques, high-stress mechanical objects
are poised to explore a new regime of exceptional quality factors.
Such quality factors combined with an optical probe at cryogenic temperatures
will have a direct impact on resonators as quantum objects, as well
as force sensors at mK temperatures. 
\end{abstract}

\date{\today}

\maketitle
Stoichiometric silicon nitride (Si$_{3}$N$_{4}$) films are a unique
material in the field of nanomechanics due to their high internal
stress that enables high mechanical quality factors along with high
resonance frequencies. Thus far, silicon-nitride based resonators
have been found to be favorable devices for studying quantum optomechanical
phenomena \citep{thompson2008strong}, precision force sensing applications
\citep{gavartin2012hybrid,Scozzaro2016JMR}, and bidirectional microwave-optical
transducers \citep{cleland2013,Andrews2014,Bagci2014Nature}. These
applications harness extremely high quality factors that are required
for observing quantum coherent effects and for precision sensing.
The key to the unique nature of membrane mechanics is that stress
provides relatively high resonance frequencies, while maintaining
low dissipation rate~\citep{unterreithmeier2010damping}.

Although numerous studies have focused on both internal and external
loss mechanisms governing the quality factors of Si$_{3}$N$_{4}$-based
resonators, the ultimate limit to the QFP has not been identified,
especially at cryogenic temperatures \citep{zwickl2008,unterreithmeier2010damping,wilson2011high,yu2012control,VengalattorePRL2014,Ghadimi2016,Tsaturyan2016}.
Specifically, a stringent upper bound for $Q$, set by thermoelastic
damping of Si$_{3}$N$_{4}$, is particularly high, reaching $3\times10^{11}$
at room-temperature~\citep{Norris01022005,zwickl2008}. Recently,
a square SiN membrane device operated at 20 mK with an electrical
readout demonstrated $Q$'s above $10^{8}$ for a mode frequency of
about 250 kHz \citep{Yuan2015}. Their observed increase in $Q$ with
decreasing temperature (below 1 K) was consistent with previous observations
of decreased internal mechanical loss in amorphous solids with temperature
\citep{Pohl2002}. Recently, it was also shown that by decreasing
the external radiation losses as well as the edge clamping losses,
room temperature Si$_{3}$N$_{4}$ can approach and surpass quality
factors of $10^{8}$ for trampoline and phononic crystal shield designs
\citep{NortePRL2016,Reinhardt2016,Tsaturyan2016}. By combining the
suppression of internal losses through cryogenic temperatures with
advanced designs for isolating the resonator to reduce external losses,
further improvements to $Q$ seem possible.

\begin{figure}[!t]
\centering{}\includegraphics[width=1\columnwidth]{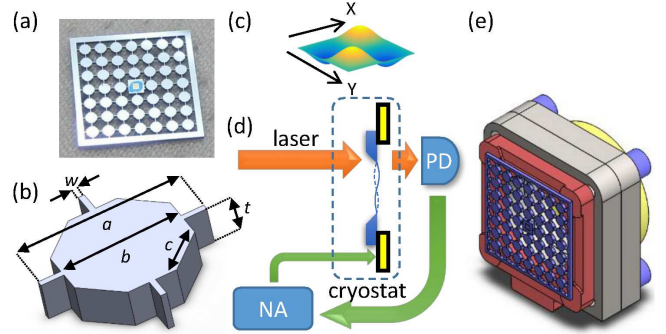} \caption{(color online) (a) A illustration of the fabricated sample composed
of phononic crystal silicon chip with the Si$_{3}$N$_{4}$ membrane
resonator at the center. The periodic structure is comprised of large
area pads connected by thin bridges. The membrane is separated from
the PnC by a membrane frame support, while the entire chip is surrounded
by a chip frame. (b) Schematic of a PnC unit cell and definitions
of the geometry parameters. See Table \ref{PnC Parameters} for the
values of these parameters. (c) Illustration of the (2,2) membrane
mode. (d) Schematic sketch of the apparatus to measure mechanical
decay time via a ringdown technique. A laser beam is transmitted through
the membrane (blue), which is mounted inside the dilution refrigerator
(dashed blue). The transmitted beam is detected by an amplified photodiode
(PD). The membrane modes are excited by a piezo ring (yellow) epoxied
onto the sample. The excitation and narrowband filtering of the optical
detection is done by a network analyzer (NA). (e) Sample assembly.
The sample (blue) is mounted on a silicon holder (red), which is mounted
on two invar steel adapters (gray). These adapters are epoxied to
the piezo actuator (yellow). The thermal link of the assembly to the
base-plate is completed with a copper foil (not-shown) pressed between
the invar adapters.}
\label{Apparatus and mode shape} 
\end{figure}

Here we study the mechanical loss of a MHz frequency Si$_{3}$N$_{4}$
membrane at temperatures down to 35 mK using piezoelectric excitation
and optical detection. We show that the mechanical loss decreases
significantly below 5 K, reaching a maximal measured $Q$ of $2.3\times10^{8}$
for a mode frequency of 1.6 MHz, corresponding to a QFP of $3.7\times10^{14}$
Hz. Additionally, the observed quality factors were sensitive to optical
probing power, further emphasizing the role of the thermal environment
in the mechanical dissipation.

The sample we utilize is a high stress (0.8 GPa) square membrane resonator
surrounded by a phononic crystal (PnC) shield. The sample is depicted
in Fig. \ref{Apparatus and mode shape}(a, b), with the parameter
values listed in Table \ref{PnC Parameters}. The devices are created
starting with a high stress Si$_{3}$N$_{4}$ film 100 nm thick grown
by low pressure chemical vapor deposition (LPCVD) on a 300 $\mu$m
silicon wafer. The final sample chip is a 10 mm x 10 mm square, with
the released Si$_{3}$N$_{4}$ membrane in the center. The main aspects
of the fabrication process are reported in \citep{yu2014phononic}.
In brief, the membrane is created through a combination of deep reactive
ion etching (DRIE), and a final KOH etch to release the membrane.
The PnC structure is created by etching all the way through the silicon
with DRIE; during the PnC creation the membrane is protected by process
adhesive that, for the devices reported here, filled the entire back
trench of the released membrane, improving yield. Further, membrane
cleanliness was improved by removing the Si$_{3}$N$_{4}$ film on
the membrane side of the device (outside of the suspended membrane
region) prior to creating the PnC. While the the device fabrication
is similar to previous work \citep{yu2014phononic}, there are several
important design changes to the PnC which should improve the robustness
and reproducibility of the band-gap, see Fig. \ref{Apparatus and mode shape}(b).
These improvements include narrowing the width of the bridges (47
$\mu$m compared to 97 $\mu$m in the previous design) to increase
the phononic isolation and designing chamfered corners to better reproduce
simulated structures in the fabrication. The design band-gap of the
PnC with the dimensions listed in Table \ref{PnC Parameters} spans
between 1.1-2.1 MHz, and between 3.1-3.3 MHz. The band-gaps were calculated
using Comsol, as described in detail in \citep{yu2014phononic}.

\begin{table}
\caption{Measured geometry parameters of the devices}
\label{PnC Parameters} %
\begin{tabular}{|c|c|c|}
\hline 
Definition  & symbol  & Device \tabularnewline
\hline 
\hline 
chip size &  & 10 mm\tabularnewline
\hline 
number of unit cells  &  & 4\tabularnewline
\hline 
unit cell size  & $a$  & 1275 $\mu$m\tabularnewline
\hline 
block length  & $b$  & 925 $\mu$m\tabularnewline
\hline 
bridge width  & $w$  & 47 $\mu$m\tabularnewline
\hline 
wafer thickness  & $t$  & 300 $\mu$m\tabularnewline
\hline 
membrane length  &  & 516 $\mu$m\tabularnewline
\hline 
chamfer size & $c$ & $\sqrt{2}\times275$ $\mu$m\tabularnewline
\hline 
membrane frame size  &  & 1125 $\mu$m\tabularnewline
\hline 
chip frame size &  & 500 $\mu$m\tabularnewline
\hline 
membrane thickness  &  & 0.1 $\mu$m\tabularnewline
\hline 
\end{tabular}
\end{table}

The experimental setup to measure the mechanical quality factor is
depicted in Fig. \ref{Apparatus and mode shape}(d, e). The sample
is anchored to the base of a dilution refrigerator, and light is directed
to the membrane through a narrow cryogenic beam path designed to filter
300 K blackbody radiation \citep{kuhn2014}. The laser beam was generated
by a diode with a 900 nm wavelength, and detected by a either a Si
avalanche or an InGaAs amplified photodiode. An interferometric signal
which monitors the position of the membrane is created due to a low
finesse cavity between the membrane and a low reflectivity mirror
inside the cryostat \citep{zwickl2008,Wilson12Thesis}. Mechanical
driving was applied by a piezoelectric ring (Noliac NAC2124). The
mechanical ringdown measurements were performed with a network analyzer
(HP 4395A), where a continuous wave excitation of the piezoelectric
ring at a specific membrane mode frequency was stopped abruptly and
the decay of the mechanical oscillation was optically monitored via
the interferometric signal. We measure the mechanical decay time constant
from the ringdown data $\tau_{m}$, and extract the mechanical loss
rate $\Gamma_{m}=\left(2\pi\tau_{m}\right)^{-1}$. The quality factor
of mode $m$, $Q_{m}$ is the ratio of the mechanical angular frequency
$\Omega_{m}$, and the loss rate $\Gamma_{m}$, $Q_{m}=\Omega_{m}/\Gamma_{m}$.

To study the temperature dependence of $Q$, we varied the base-plate
temperature of the refrigerator from 35 mK to 5 K. The base-plate
temperature was measured with a calibrated RuO$_{2}$ resistive thermometer.
The thermal link of the membrane to the base-plate is achieved with
a copper foil pressed between the invar adapters, shown in Fig. \ref{Apparatus and mode shape}(e).
The thermalization efficiency down to 120 mK was confirmed by an independent
measurement performed on the same setup and sample, utilizing a 1064
nm resonant high finesse cavity to probe the thermally-induced brownian
motion of a membrane. There we obtained the thermal phonon occupancy
via optomechanical relations \citep{peterson2016laser} and found
good agreement with the base-plate temperature down to 120 mK \citep{Kampel2016}.

\begin{figure}[tb!]
\begin{centering}
\includegraphics[width=1\columnwidth]{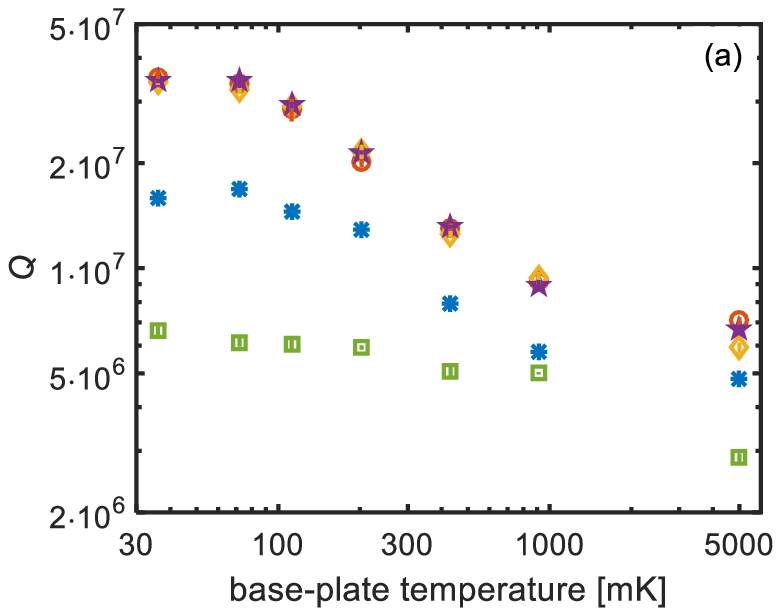} 
\par\end{centering}

\begin{centering}
\includegraphics{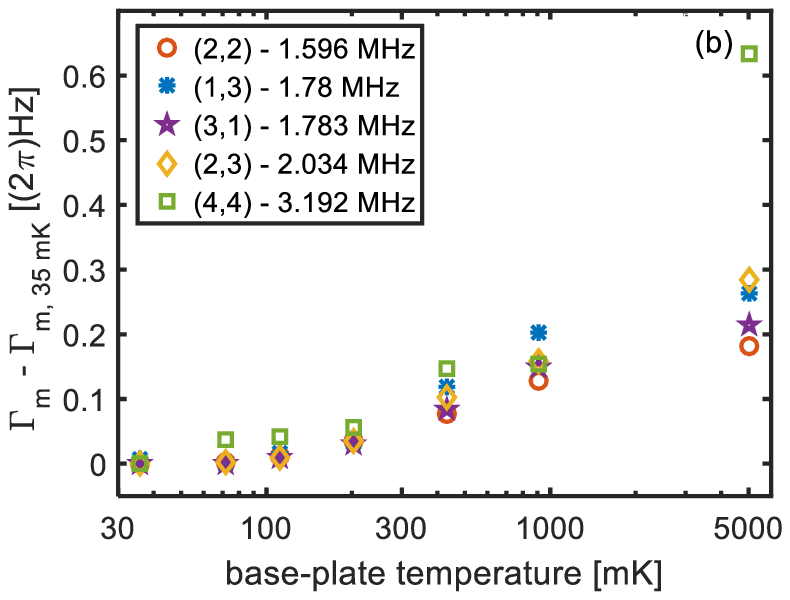} 
\par\end{centering}

\centering{}\caption{Measurements of $Q$-factor (a) and difference in loss rate from minimal
loss at 35 mK (b) of multiple membrane modes versus base-plate temperature
and at an optical power 1 nW. Errorbars are standard deviation of
repeated measurements.}
\label{Temperature Scan} 
\end{figure}

At each temperature, we measured the mechanical $Q$ of a few membrane
modes that displayed values higher than $10^{6}$. These mode frequencies
were inside the design band-gap of the PnC. Other modes that were
outside of the band-gap of the PnC displayed values lower than $10^{5}$.
As expected, the decrease in the base-plate temperature resulted in
an increase in the value of $Q$ (Fig. \ref{Temperature Scan}(a)),
and correspondingly a decrease in the loss rate $\Gamma_{m}$ (Fig.
\ref{Temperature Scan}(b)). In these data, we witness a leveling-off
of the $Q$-values at the low end of the temperature range. The leveling-off
of the $Q$ between 200 mK to 1 K, does not occur here, in contrast
to the results in \citep{Yuan2015}, and continues to drop above 1
K, similar to the results of \citep{zwickl2008}. To make this feature
of the data more apparent, we plot the difference in the loss rate
$\Gamma_{m}$ from that at 35 mK in Fig. \ref{Temperature Scan}(b).
Notice that this quantity increases monotonically above 100 mK

\begin{figure}[tb!]
\centering{}\includegraphics[width=1\columnwidth]{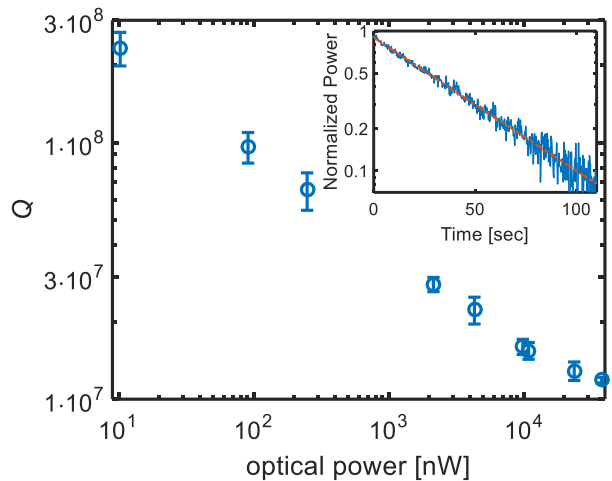} \caption{Mechanical ringdown results at a base-plate temperature of 35 mK versus
input laser power measured via free-space optical probe at 900 nm.
The errorbars are calculated from repeated measurements at the same
laser power. Inset: Mechanical ringdown of mode (2,2) at a laser power
of 10 nW (blue), and an exponential fit with a decay rate of $\Gamma_{m}=\left(2\pi\right)7.05\pm0.005$
mHz. Errorbars are standard deviation of repeated measurements.}
\label{Laser Power Scan} 
\end{figure}

For the above measurements, we used a free space optical probe near
900 nm, operated continuously. By varying the probe's optical power
we examined its effect on the resonator's dissipation. As we increase
the probe's optical power, we find that the mechanical quality factor
decreases significantly (Fig. \ref{Laser Power Scan}), presumably
due to heating resulting from optical absorption. While we do expect
that heating effects become more important with the decrease of thermal
conductivity of amorphous solids below 1 K \citep{Leivo1998,Pohl2002,Wlison2015},
we stress that the measurements here do not represent a quantitative
study of the optical properties of Si$_{3}$N$_{4}$ itself at cryogenic
temperatures; for example, we have not characterized the spatial mode
of the optical spot on the membrane. We note the effective power of
a 1064 nm beam, probing an optical cavity displaying heating effects,
was measured to be considerably higher than for the 900 nm optical
probe power for which we see an effect here. Specifically, previous
work shows no heating effects at 120 mK \citep{Kampel2016}, and a
heat transfer analysis at 35 mK indicates much smaller optical effects
than those shown in Fig. \ref{Laser Power Scan} would cause significant
effects to the Si$_{3}$N$_{4}$ alone. Further, we note that optical
excitation effect on mechanical quality factor is a complex subject,
which cannot be simply described by a heating of a thermal bath \citep{KrausePRL2015}.
With the data at hand, a complete picture cannot be discerned. The
study of optical properties of Si$_{3}$N$_{4}$ membranes will be
the subject of a future study.

Nonetheless, the $Q$ dependence on optical absorption is indicative
of variation of an internal dissipation as a function of temperature.
Salient features we observe are that at a minimal probe power of 10
nW, and temperature of 35 mK, we find a maximal measured $Q$ of $2.3\pm0.3\times10^{8}$
for the (2,2) mode, corresponding to a QFP of $3.7\times10^{14}$
Hz as seen in Fig. \ref{Laser Power Scan}. The error bar shown is
dominated by the low signal to noise level at the light level of the
probe. We note the $Q$ values of Fig. \ref{Laser Power Scan} are
higher than the values depicted in Fig. \ref{Temperature Scan} and
were obtained prior to thermal cycling of the refrigerator. We suspect
that the thermal cycling resulted in an added loss that obscured the
reduction in dissipation below 100 mK in the data of Fig. \ref{Temperature Scan}.

In conclusion, we optically probed the mechanical loss of MHz frequency
Si$_{3}$N$_{4}$ membranes down to mK temperatures. The highest $Q$
we measured was $2.3\times10^{8}$ at 35 mK, corresponding to a QFP
of $3.7\times10^{14}$ Hz, the highest value achieved for Si$_{3}$N$_{4}$
resonators so far. These results push further the performance of these
resonators at cryogenic temperatures, both for quantum studies, as
well as for precision measurements. A next step would be to combine
cryogenic low-loss membrane resonator with additional geometric design
features aiming to further reduce internal and external loss channels
to reach even higher levels of isolation. Finally, the low thermal
conductivity and the high-$Q$ of Si$_{3}$N$_{4}$ membranes at mK
temperatures allows optical heating effects to be measured. This enables
a quantitative study of optical properties of Si$_{3}$N$_{4}$ membranes.

We thank A. Higginbotham for technical assistance. This work was supported
by AFOSR PECASE, ONR DURIP, AFOSR-MURI, Rafael Ltd., the Cottrell
Scholar\textquoteright s program, and the National Science Foundation
under grant number 1125844.

Contributions from NIST are not subject to U.S. copyright. Reference
to specific products and services does not constitute endorsement
by NIST. Other vendors may provide equivalent or better products.

\bibliographystyle{apsrev4-1}
\bibliography{highQ_RF7}

\end{document}